\documentclass[twocolumn]{aastex701}
 
\usepackage[dvipsnames]{xcolor}
\definecolor{linkcolor}{rgb}{0.1, 0.5, 0.7}

\hypersetup{
    colorlinks=true,
    linkcolor=linkcolor,
    citecolor=linkcolor,
    filecolor=linkcolor,
    urlcolor=linkcolor,
}
 
\usepackage{amssymb}
\usepackage{amsmath}
\usepackage{physics}
\usepackage{acronym}
\usepackage{booktabs}
\usepackage[absolute,overlay]{textpos}
\usepackage[normalem]{ulem}

\usepackage{eso-pic}

\newcommand{\msun}{\ensuremath{\mathrm{M}_\odot}}
 
\graphicspath{{./figures}{./figures/binned}{./figures/injection}{./figures/subpop}}

\newacro{LVK}[LVK]{LIGO--Virgo--KAGRA}
 
\newacro{BBH}{binary black hole}
\newacro{GW}{gravitational wave}
\newacro{SNR}{signal-to-noise ratio}
\newacro{FAR}{false-alarm rate}
\newacro{KL}{Kullback--Leibler}
\newacro{HMC}{Hamiltonian Monte Carlo}
\newacro{VI}{variational inference}
\newacro{PSIS}{Pareto-smoothed importance sampling}
\newacro{ICAR}{intrinsic conditional autoregressive}
\newacro{PPD}{posterior population distribution}
\newacro{XG}{next-generation}
\newacro{BH}{black hole}
\newacro{GWTC}{gravitational-wave transient catalog}
 
\begin{document}

\AddToShipoutPictureFG*{%
  \AtPageUpperLeft{%
    \put(35,-115){\includegraphics[width=4cm]{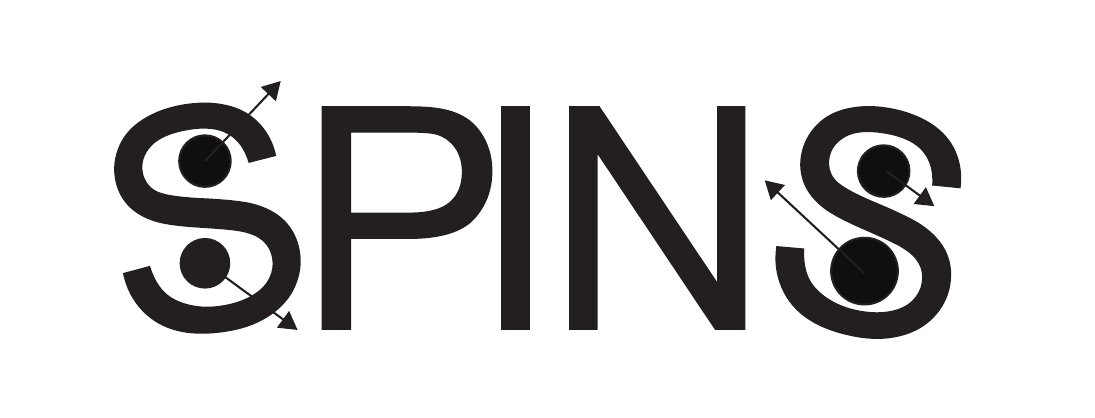}}%
  }%
}
 
\shorttitle{No evidence for a peak in the BBH spin tilt distribution}
\shortauthors{Wolfe et al.}
 
\title{No model-independent evidence for a peak in binary black hole spin (mis)alignments}

\author[0000-0003-2540-3845]{Noah E. Wolfe}
\affiliation{LIGO Laboratory, Massachusetts Institute of Technology, Cambridge, MA 02139, USA}
\affiliation{Kavli Institute for Astrophysics and Space Research, Massachusetts Institute of Technology, Cambridge, MA 02139, USA}
\affiliation{Department of Physics, Massachusetts Institute of Technology, Cambridge, MA 02139, USA}
\email[show]{newolfe@mit.edu}
 
\author[0000-0003-2700-0767]{Salvatore Vitale}
\affiliation{LIGO Laboratory, Massachusetts Institute of Technology, Cambridge, MA 02139, USA}
\affiliation{Kavli Institute for Astrophysics and Space Research, Massachusetts Institute of Technology, Cambridge, MA 02139, USA}
\affiliation{Department of Physics, Massachusetts Institute of Technology, Cambridge, MA 02139, USA}
\email{svitale@mit.edu}
 
\author[0000-0002-0147-0835]{Michael Zevin}
\affiliation{Adler Planetarium, 1300 South DuSable Lake Shore Drive, Chicago, IL 60605, USA}
\affiliation{Center for Interdisciplinary Exploration and Research in Astrophysics (CIERA), Northwestern University, 1800 Sherman Ave, Evanston, IL 60201, USA}
\email{mzevin@adlerplanetarium.org}
 
\collaboration{all}{(Society of Physicists Interested in Non-aligned Spins, SPINS)\footnote{\url{www.sites.mit.edu/spins}}}
\noaffiliation{}

\begin{abstract}

The degree of black-hole spin-orbit misalignment (``tilts'') in the astrophysical population could be a powerful diagnostic to distinguish between binary formation in isolation, in dynamical environments, or in hierarchical triples.
However, robust population-level spin tilt measurements are complicated by model misspecification as well as numerical and Poisson variance, ultimately owing to poor single-event constraints on tilts.
Motivated by reports of a possible peak in the spin tilt distribution,
we analyze the fourth LIGO-Virgo-KAGRA gravitational-wave transient catalog
with two population models designed
to test for preferred spin orientations at different black hole masses.
We find that a peak in spin tilts is neither statistically significant nor model independent.
Since the data cannot be used to reliably identify subpopulations based on their spin tilt properties, we also consider a complementary approach: measuring the spin magnitude and tilt distributions at fixed mass scales. We find no confident correlation between mass and spin tilt, but we do confirm a confident correlation between spin magnitude and mass, corroborating recent analyses.

\end{abstract}

\section{Introduction} \label{sec:intro}

Gravitational-wave (\acs{GW})\acused{GW} observations of merging \acp{BBH} have quickly become routine since their discovery in 2015;
the latest public \ac{LVK} \citep{LIGOScientific:2014pky, VIRGO:2014yos, KAGRA:2020tym} catalog now stands at 204 candidates \citep{LIGOScientific:2025slb}.
Together, these data can be used to infer the population distribution of black holes in merging binaries \citep{LIGOScientific:2025hdt, LIGOScientific:2025yae, LIGOScientific:2025pvj}
which may reveal their formation channels.
Black-hole binaries are characterized, in part, by their masses and spin vectors.
In particular, the orientation---the tilt---of their spins relative to the binary orbit
may be a clear tracer of \ac{BBH} formation histories.
For example, simulations of binary evolution in the field typically predict \acp{BBH}
with orbit-aligned spin tilts \citep{Kalogera:1999tq, Farr:2017uvj, Gerosa:2018wbw, Belczynski:2017gds, Baibhav:2024rkn}.
Meanwhile, models of binary formation in stellar clusters typically predict
an isotropic distribution of spin tilts \citep{2016ApJ...832L...2R, 2022ApJS..258...22R}.

Black hole (\acs{BH})\acused{BH} spin orientations may be even more astrophysically
interpretable if we can associate different tilt distributions with different mass scales,
e.g., field formation may tend to produce \acp{BH} below the onset of pulsational
pair-instability at $\sim 45\,\msun$ \citep{Farmer:2019jed}
while formation in clusters may yield heavier \acp{BH} \citep{ Gerosa:2021mno, 2022ApJS..258...22R}.
Merging \acp{BBH} may also be formed in hierarchical triples,
where the inner \ac{BH} binary is driven to merger via dynamical interactions with an outer object.
The coupled dynamics of two \ac{BH} orbits and spins alongside a third body,
their gravitational-wave emission,
and Kozai-Lidov oscillations leave a unique observational imprint:
the inner binary components tend to spin in the orbital plane
\citep{Antonini:2017tgo, Rodriguez:2018jqu, Liu:2018nrf, Yu:2020iqj, Su:2020vda}.

Analyses of gravitational-wave catalogs
have found hints of \ac{BH} mass--spin correlations,
e.g., \citealt{Callister:2021fpo, Li:2023yyt, Godfrey:2023oxb, Pierra:2024fbl}.
Recently, the \acs{LVK}'s fourth gravitational-wave transient catalog (GWTC-4.0) 
has provided additional evidence of complicated structure in the joint space of \ac{BH} masses and spins \citep{LIGOScientific:2025pvj, Banagiri:2025dmy, Antonini:2025ilj, Guttman:2025jkv, Berti:2025usa, Tong:2025xir, Li:2025iux, Vijaykumar:2026zjy, Farah:2026jlc, Plunkett:2026pxt, Ray:2026uur, Galaudage:2026opk}.
The \acs{LVK}'s analysis of GWTC-4.0 allows for a preference for in-plane spin tilts \citep{LIGOScientific:2025pvj}.
Further, recent work by \citealt{Stegmann:2025zkb} associates this possible peak in the spin tilt distribution with a subpopulation of \acp{BBH} below $\sim 45\,\msun$,
which could be a signature of hierarchical triples formed in the field.
Their analysis mildly prefers low-mass \acp{BH} to have in-plane vs. orbit-aligned spin tilts and suggests that hierarchical triples are the dominant channel through which merging black-hole binaries form.
However, their model assumes a less flexible form for the primary mass distribution than the parametric \ac{LVK} analysis.

While spin tilts are interpretively powerful---and current gravitational-wave observations may even suggest that \acp{BH} at different mass scales tend to prefer certain spin orientations---it is challenging to measure \ac{BBH} spin tilts via \acp{GW}.
Instead, we typically measure best the effective spin of the binary  \citep{Purrer:2015nkh, Vitale:2016avz, Shaik:2019dym, Pratten:2020igi, Green:2020ptm, Biscoveanu:2021nvg, Krishnendu:2021cyi, Miller:2025eak}, i.e.,
the mass-weighted projection
of the component spin vectors along the orbital angular momentum \citep{Damour:2001tu, Racine:2008qv, 2014LRR....17....2B}.
In principle,
population inference should be able to make a meaningful
measurement of spin tilts from the combination of many poorly measured examples \citep{Vitale:2015tea, Talbot:2017yur}.
In practice and given current \ac{GW} datasets,
population constraints on spin tilts are:
sensitive to modeling assumptions\footnote{
Population inference of spin tilts may also be sensitive to choice of waveform model at the single-event level,
particularly for highly-spinning \acp{BBH}
\citep{2025PhRvX..15c1036D}.
} \citep{Vitale:2022dpa, Golomb:2022bon, Miller:2024sui};
sensitive to Poisson fluctuations in catalog membership \citep{Vitale:2025lms, Corelli:2026thw};
and limited by systematic uncertainties in the population likelihood \citep{Talbot:2023pex, Heinzel:2025ogf}.
While some population analyses allow (but do not require) a preference for \acp{BH} with in-plane tilts,
these inferences may be spurious.

Owing to the challenges involved in making population-level spin tilt measurements,
here we further investigate the
possibility of a preferred spin orientation
in the \ac{BBH} population,
while allowing for subpopulations or correlations in the joint distribution of \ac{BBH} masses and spin tilts.
In Sec.~\ref{sec:methods} we describe our population models and inference methods,
in Sec.~\ref{sec:results} we detail the inferred spin tilt distribution under each model,
and in Sec.~\ref{sec:discussion} we summarize our findings.

\section{Methods} \label{sec:methods}

\subsection{Hierarchical inference} \label{sec:inference}

We infer the population-level properties of merging \acp{BBH} via hierarchical Bayesian inference \citep{Loredo:2001rx, 2019MNRAS.486.1086M, Vitale:2020aaz}.
We take the confident (signal-to-noise ratio $> 10$ through the \acs{LVK}'s second observing run and false-alarm rate $< 1\,\mathrm{yr}^{-1}$ thereafter) \acp{BBH} in GWTC-4.0 \citep{ligo_scientific_collaboration_and_virgo_2025_16053484}.
For each event in O3 and O4a, we use source parameter inference assuming the \textsc{NRSur7dq4} \citep{Varma:2019csw} waveform model when available and the \textsc{Mixed} samples otherwise;\footnote{See Sec.~3 of \citealt{LIGOScientific:2025pvj} for details on \textsc{Mixed} sample sets.}
in O1 and O2 we use \textsc{IMRPhenomXPHM} \citep{Pratten:2020ceb, Colleoni:2024knd}.
Selection effects are estimated with publicly-available \ac{LVK} data products \citep{Essick:2025zed, o4a-sensitivity-zenodo, gwtc4-cumulative-sensitivity-zenodo}.
Event selection, waveform choices, and sensitivity estimation follow \citealt{LIGOScientific:2025pvj}.
We estimate the population likelihood via Monte Carlo integration
and limit the variance of this estimator to be less than one during inference
\citep{Tiwari:2017ndi, Essick:2022ojx, Talbot:2023pex, Heinzel:2025ogf}.
Inference is performed using \texttt{dynesty} \citep{2020MNRAS.493.3132S, sergey_koposov_2024_12537467} wrapped by
\texttt{gwpopulation} \citep{Talbot2025};
settings are detailed in App.~\ref{app:sampling}.

\subsection{Population models} \label{sec:models}

We construct two population models designed to test for the existence of subpopulations or correlations in the joint space of black hole masses and spins,
which we call the \textsc{Subpopulations} and \textsc{Correlation} models, respectively.
Both models and their priors are detailed in App.~\ref{app:models}.
We also designed more elastic models,
although their posterior geometries were non-trivial to sample
so we reserve description of these model variations and sampling challenges for App.~\ref{app:sampling}.

\textsc{Subpopulations}:
To model the astrophysical \ac{BBH} distribution,
we follow
the mixture model presented in \citealt{Stegmann:2025zkb} for \ac{BBH} spins
while adopting a model for \ac{BBH} masses which is consistent with the \ac{LVK}'s analysis of GWTC-4.0 \citep{LIGOScientific:2025pvj}.
In detail, for mass and redshift we adopt the preferred parametric distribution from \citealt{LIGOScientific:2025pvj}:
we model the primary masses as a mixture between a broken power law tapered at low masses and two Gaussian peaks
(\textsc{Broken Power law plus Two Peaks}),
we model mass ratios as a power law,
and we model the differential merger rate density $\mathcal{R}$ as a power law in redshift.
For simplicity, we use the same minimum \ac{BH} mass for primary and secondary components.
The spin tilts follow \citealt{Stegmann:2025zkb},
with a model that identifies \ac{BBH} subpopulations based on spin tilt and primary mass.
This model is inspired by two potential astrophysical formation channels for merging \acp{BBH}:
those formed in field binaries or triples which may prefer some spin orientation,
and \acp{BBH} formed via dynamical interactions in dense environments which may have an isotropic spin distribution.
This model accounts for a possible maximum mass for field \acp{BH} (e.g., the onset of pulsational pair-instability)
by transitioning from a mixture of tilt distributions to isotropic only above a certain mass.
Specifically, for \acp{BH} with primary mass above a transition mass $\tilde{m}$ (dubbed the ``high-mass'' region)
cosine-spin tilts are uniformly distributed,
while below $\tilde{m}$ (the ``low-mass'' region) they are a mixture between a uniform distribution and a truncated Gaussian.
The transition mass, Gaussian location and scale, and branching ratio between the Gaussian and uniform components are free parameters.
We use the same prior for the scale of the truncated Gaussian in spin tilts as the preferred analysis in \citet{Stegmann:2025zkb}.
Each subpopulation (low-mass + isotropic; low-mass + Gaussian; high-mass + isotropic) has its own spin magnitude distribution parameterized as a truncated Gaussian.
The spin magnitude and tilt distributions for each subpopulation are identically distributed for primary and secondary components.

\textsc{Correlation}:
Instead of distinguishing features in the population based on their spin tilts,
we attempt to measure the distribution of \ac{BH} spins at different mass scales
by coarsely binning their masses and defining different spin magnitude and tilt distributions in each bin.
We model each \ac{BH} component spin magnitude $\chi$ with a truncated Gaussian,
\begin{equation} \label{eq:correlation-spin-mag}
    p(\chi \mid m) = \mathcal{N}_{[0, 1]}(\chi \mid \mu_\chi(m), \sigma_\chi(m)) \, ,
\end{equation}
and each component cosine-spin tilt $\cos \tau$ with a mixture between uniform and truncated Gaussian distributions,
\begin{equation} \label{eq:correlation-spin-tilt}
\begin{split}
    p(\cos \tau \mid m) = \big[ &\xi(m) \mathcal{N}_{[-1, 1]}(\cos \tau \mid \mu_\tau(m), \sigma_\tau(m)) \\
    &+ \frac{1}{2} (1- \xi) \big] \, , \\
\end{split}
\end{equation}
where $\mathcal{N}_{[a,b]}(\cdot \mid \mu, \sigma)$
denotes a truncated Gaussian distribution on a domain $[a, b]$ with location $\mu$ and scale $\sigma$. Here, the location ($\mu_\chi, \mu_\tau$),
scale ($\sigma_\chi, \sigma_\tau)$,
and branching ratio $\xi$ parameters depend on \ac{BH} mass $m$.
These parameters only depend on $m$ and are not conditioned on primary vs. secondary component.
The spin population parameters are determined by which of four bins the \ac{BH} mass falls in: $[3, 7],\,[7, 20],\,[20, 40],\,$ and $[40, 300]\,\msun{}$.
These bins are chosen to straddle known features in the primary mass distribution at $\sim 10\,\msun$ and $\sim 35\,\msun$ \citep{LIGOScientific:2025pvj},
as well as possible transitions in the spin distribution \citep{Tong:2025xir, Plunkett:2026pxt}.
Note that the spins of the primary and secondary components are \textit{not} identically distributed
as the primary and secondary masses are not identically distributed;
instead, the spins of black holes \textit{falling in the same mass bin} are identically distributed.
As in the \textsc{Subpopulations} model,
we assume a power law in the differential merger rate density over redshift, a \textsc{Broken Power law Plus Two Peaks} in primary mass, and a power law in mass ratio.
Results with an alternative choice of bins are presented in App.~\ref{app:altbins}
and lead to similar conclusions detailed hereafter.

\begin{figure*}
    \centering
    \includegraphics[width=0.845\linewidth]{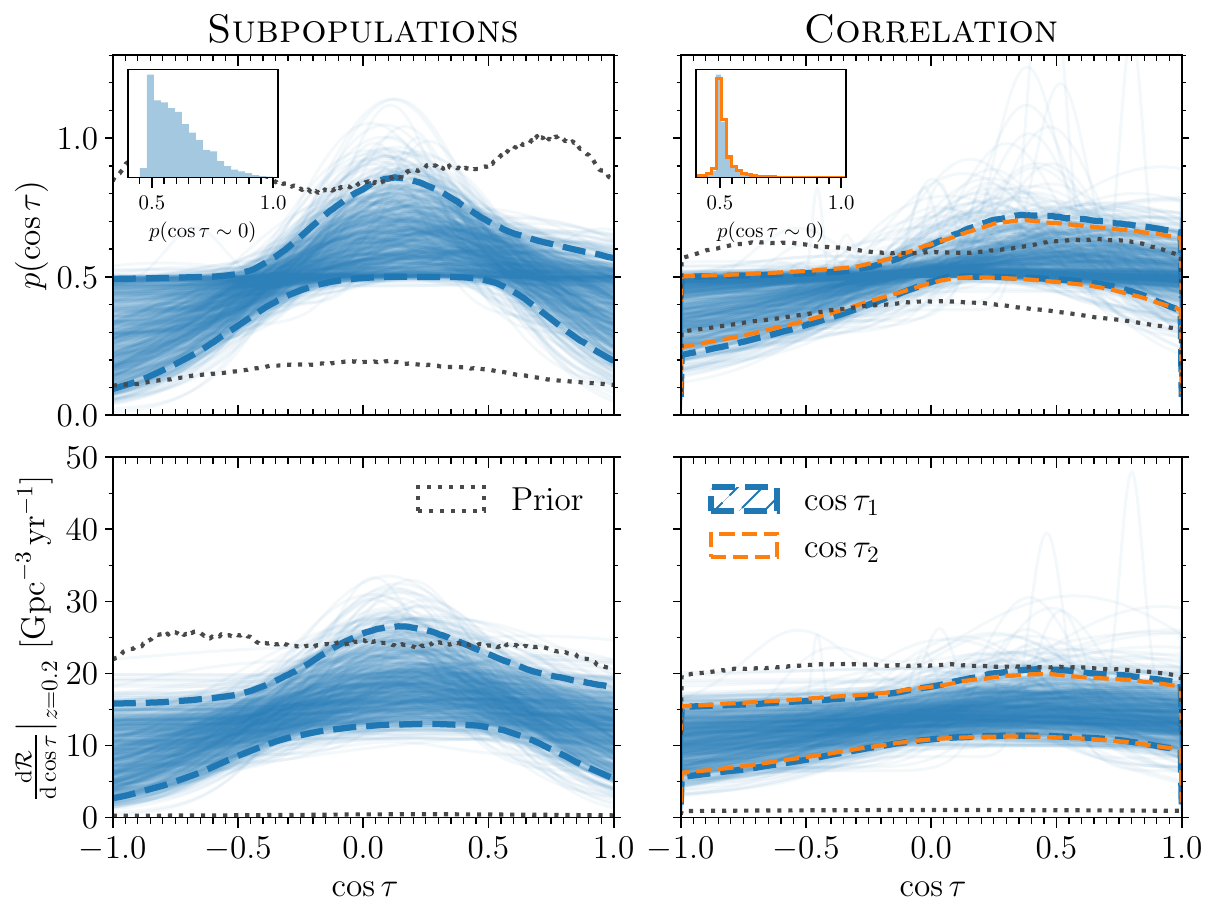}
    \caption{
        \acsp{PPD} of the marginal density (top)
        and differential merger rate (bottom) of sources as a function of
        spin tilt $\tau$ under the \textsc{Subpopulations} model (left)
        and \textsc{Correlation} model (right).
        Dashed lines enclose the 90\% credible region and transparent lines
        are $10^3$ random draws from the posterior.
        Since component tilts are not identically distributed in the
        \textsc{Correlation} model, we show the 90\% credible bounds for
        both the primary (blue) and secondary (orange).
        In the top row,
        we include insets with the posterior on the
        marginal population density near $\cos \tau \sim 0$.
        We show the 90\% credible bounds of the prior population distributions with gray dotted lines;
        insets are reproduced with priors in Fig.~\ref{fig:inset-with-prior}.
    } 
    \label{fig:tilts-total-ppd}
\end{figure*}

\begin{figure}
    \centering
    \includegraphics[width=0.85\linewidth]{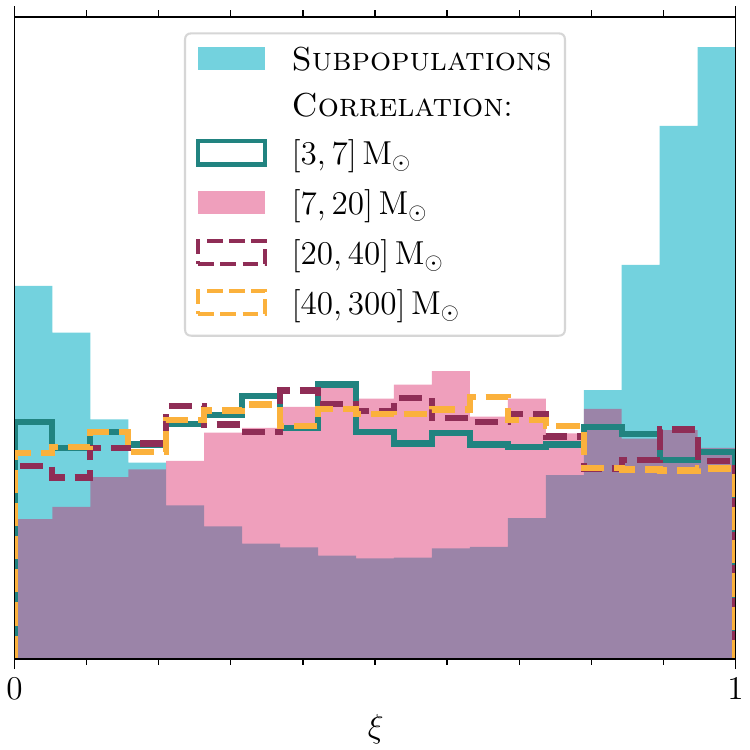}
    \caption{
        The fraction $\xi$ of sources in the Gaussian peak below $\tilde{m}$ in the \textsc{Subpopulations} model (blue),
        as well as $\xi$ within each mass bin under the \textsc{Correlation} model.
        }
    \label{fig:xi}
\end{figure}

\section{Results} \label{sec:results}

In this section, we present posteriors on the population-level spin distributions obtained under both models.
For completeness,
we show marginal posteriors on the parameters of the redshift and mass distributions in Fig.~\ref{fig:mass-redshift}.
Throughout, we refer to 90\% symmetric credible intervals unless otherwise stated.

\subsection{Marginal tilt distributions}

In Fig.~\ref{fig:tilts-total-ppd} we show the \acp{PPD} of the marginal density
and differential merger rate as a function of spin tilt recovered
by both model variations.
We show the differential merger rate at $z = 0.2$, consistent with the \ac{LVK}'s population analysis of GWTC-4.0 \citep{LIGOScientific:2025pvj}.
We see that under the \textsc{Subpopulations} model, a preference for in-plane
($\cos \tau \sim 0$) tilts is allowed within the 90\% credible region but not required.
We can quantify the lack of a statistically-significant preference for in-plane spin tilts
following \citealt{Vitale:2022dpa} and \citealt{Stegmann:2025zkb}:
we compute the relative fraction\footnote{See also \citealt{Vitale:2025lms} which demonstrated that integrated measurements more robustly quantify structure in the tilt distribution than the population model parameters.} $Y$ of sources with spin tilts 
in windows with a width of 0.1 around
$\cos \tau = 0$ vs. $\cos \tau = 1$,
\begin{align}
    Y = \frac{p(-0.05 \leq \cos \tau \leq 0.05)}{p(0.9 \leq \cos \tau \leq 1)} \, .
\end{align}
When $Y \gg 1$ ($Y \ll 1$), many more \acp{BH} have in-plane (orbit-aligned) spins as opposed to orbit-aligned (in-plane) spins; $Y = 1$ indicates no preference between these spin orientations.
We find $Y = 1.3^{+2.3}_{-0.4}$ (median and 90\% credible bounds) for the \textsc{Subpopulations} model.

Of course, $Y$ alone does not explicitly capture all possible structure in the population distribution, particularly below $\cos \tau \sim 0$;
further, for tilt distributions that approach isotropy,
$Y$ will be sensitive to the window used to define spin tilts as ``in-plane'' or ``orbit-aligned''.
However, we do not recover any confident structure in the \ac{PPD} (cf. Fig.~\ref{fig:tilts-total-ppd});
instead, we find a mode of nearly flat marginal densities in tilt (darker region indicating higher posterior density).
We highlight the inferred shape of the population distribution around $\cos \tau \sim 0$
with the inset in the top-left panel of Fig.~\ref{fig:tilts-total-ppd}.
In detail, we show the posterior on the marginal density of the population distribution of spin tilts;
to capture possible structure near $\cos \tau = 0$
we report the marginal population density averaged over a window of $-0.1 \leq \cos \tau \leq 0.1$.
Isotropic tilt distributions will have $p(\cos \tau \sim 0) = 0.5$,
whereas those with a peak at $\cos \tau \sim 0$ will have $p(\cos \tau \sim 0) > 0.5$.
We find the strongest posterior support for an isotropic distribution of spin tilts,
and less support for a distribution which peaks near $\cos \tau \sim 0$.
This is reflected in Fig.~\ref{fig:xi}, where we show the marginal posterior
on the fraction $\xi$ of sources
in the Gaussian peak inferred under both the \textsc{Subpopulations} and \textsc{Correlation} models.
For the \textsc{Subpopulations} model, we see modes at both $\xi \sim 0$ and $1$.
These modes occur with similar likelihood;
computing log-likelihoods relative to the highest likelihood point found with this model,
posterior samples with $\xi < 0.2$
have log-likelihood $-9.9^{+4.0}_{-5.8}$
while samples with $\xi > 0.8$ have log-likelihood $-7.8^{+4.0}_{-6.0}$.
We note that these likelihoods are consistent within statistical uncertainty.
Further, this difference in log-likelihoods ($\sim 2$) is comparable to the Monte Carlo variance of the log-likelihood estimator ($\sim 1$; cf. Fig.~\ref{fig:mutau-sigtau-xitau}).

We inspect \acp{PPD} in primary and secondary-component tilts
under the \textsc{Correlation} model in the right column of Fig.~\ref{fig:tilts-total-ppd}.
We find that the tilt distributions for either \ac{BBH} component may be approximately isotropic
or have a broad (width of $\sim 1$) peak at $\cos \tau > 0$
(see also Fig.~\ref{fig:mutau-sigtau-xitau}).
Considering the inset in the top-right panel of Fig.~\ref{fig:tilts-total-ppd},
the data prefer the primary and secondary-component tilt distributions to have marginal densities of $\sim 0.5$ near $\cos \tau \sim 0$,
consistent with an isotropic distribution.
Further, we find that the fraction of sources with in-plane vs. orbit-aligned spin tilts
is $Y = 0.98^{+0.5}_{-0.2}$
($0.98^{+0.4}_{-0.2}$) for primary (secondary) black holes.
While the data allow for a slight preference of in-plane vs. aligned tilts,
they disfavor any more than a $\sim 50\%$ relative excess ($Y \sim 1.5$) of \acp{BH} with in-plane over orbit-aligned spins at $\cos \tau \sim 0$.
That both primary- and secondary-component spin tilt distributions are consistent
with isotropy is reflected in Fig.~\ref{fig:xi};
there, we see that, in all mass bins, $\xi$ cannot be constrained away from
zero and is essentially unmeasured.

From Fig.~\ref{fig:xi} we note that $\xi$ has stronger support at the edges of its domain under the \textsc{Subpopulations} model vs. the \textsc{Correlation} model.\footnote{
Testing that this is not a prior effect (cf. \citealt{Biscoveanu:2025jpc}),
we found consistent results with a uniform prior on the width of the peak in tilts as opposed to a truncated Gaussian prior (cf. Tab.~\ref{tab:spin-priors}).}
However, in the \textsc{Subpopulations} case,
we model the spins of low-mass black holes---their tilts \textit{and} magnitudes---as a mixture with branching ratio $\xi$ (see Eqn.~3 in \citealt{Stegmann:2025zkb}).
Thus, for this model, the posterior on $\xi$ is informed by the spin magnitude distribution.
In contrast, the \textsc{Correlation} analysis does not jointly model magnitude and tilts, and hence $\xi$ is only informed by the latter.
Therefore, Fig.~\ref{fig:xi} tells us that: (i) under the \textsc{Subpopulations} model,
the data
do not require
a mixture of spin magnitude--tilt distributions at low masses, since there is a mode at $\xi \sim 0$,
and (ii) under the \textsc{Correlation} model the mixing fraction of tilt distributions is poorly constrained at all masses.
See also Fig.~\ref{fig:mutau-sigtau-xitau} where we record joint marginal posteriors on the parameters of the spin tilt distribution obtained under both the \textsc{Subpopulations} and \textsc{Correlation} models.

\subsection{Spin (mis)alignment-mass correlations} \label{sec:corr}

\begin{figure}
    \centering
    \includegraphics[width=0.9\linewidth]{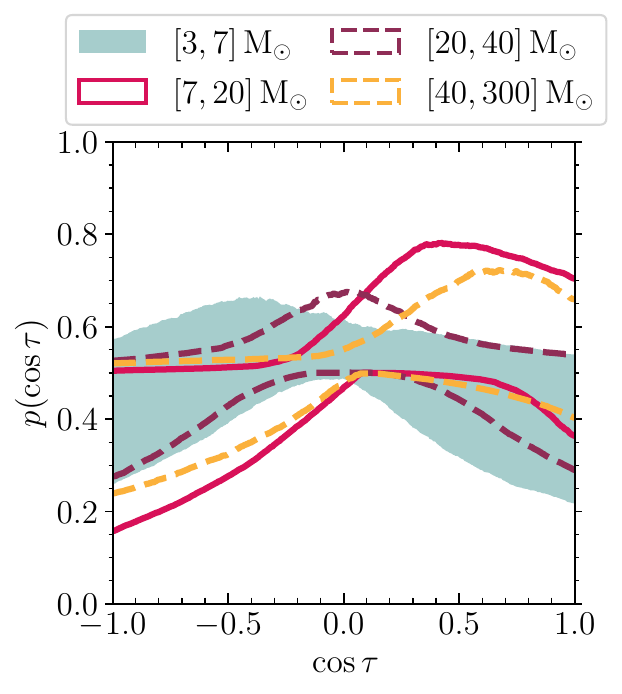}
    \caption{
        \acp{PPD} on the marginal density of sources as a function of spin tilt
        $\cos \tau$ in four bins of \ac{BH} mass under the \textsc{Correlation} model.
    }
    \label{fig:tau-per-bin}
\end{figure}

Under the \textsc{Subpopulations} model,
we infer a transition mass of $\tilde{m} = 45.1^{+15.7}_{-7.0}\,\msun$.
Below $\tilde{m}$,
the relative fraction of in-plane vs. orbit-aligned spins is $Y = 1.3^{+2.4}_{-0.4}$.
This measurement is essentially the same as with the entire population, as lighter \acp{BH} tend to dominate the inferred astrophysical population.
Since we assume an isotropic tilt distribution above $\tilde{m}$,
there is, by definition, no preference for any spin orientation in the low- or high-mass subpopulations, and so the data do not require $Y$ to evolve with mass.
This is a myopic reflection of the overall tilt distribution above and below $\tilde{m}$;
recalling Fig.~\ref{fig:xi}, tilts below the transition are consistent
with an isotropic distribution and thus with the tilt distribution at higher mass.

In Fig.~\ref{fig:tau-per-bin},
we show \acp{PPD} in the marginal density over spin tilt within each mass bin
under the \textsc{Correlation} model.
At the 90\% credible level, there is no evidence for a correlation between
mass and spin tilt.
While it is tempting to see a ``peak'' in the marginal density between
$\cos \tau \sim 0.5$--$1$ in the $[7, 20]$ and $[40, 300]\,\msun$ bins,
we emphasize that the population posteriors overlap
within the 90\% credible interval at all masses.
In particular,
spin tilts in all mass bins remain consistent
with isotropic or nearly-isotropic distributions.
Additionally, across all mass bins we find no statistically-significant preference for \acp{BH} to spin in-plane vs. aligned with their orbit;
computing the relative fraction $Y$ of \acp{BH} with in-plane vs. orbit-aligned tilts, we find $Y = 1.0^{+1.5}_{-0.1},\,1.0^{+0.6}_{-0.3},\,1.0^{+1.2}_{-0.1}$ and $1.0^{+0.3}_{-0.3}$
for \acp{BH} with masses $[3, 7],\,[7, 20],\,[20, 40]$ and $[40, 300]\,\msun{}$, respectively.

\subsection{Spin magnitude-mass correlations} \label{sec:spin-mag}

\begin{figure}
    \centering
    \includegraphics[width=0.8\linewidth]{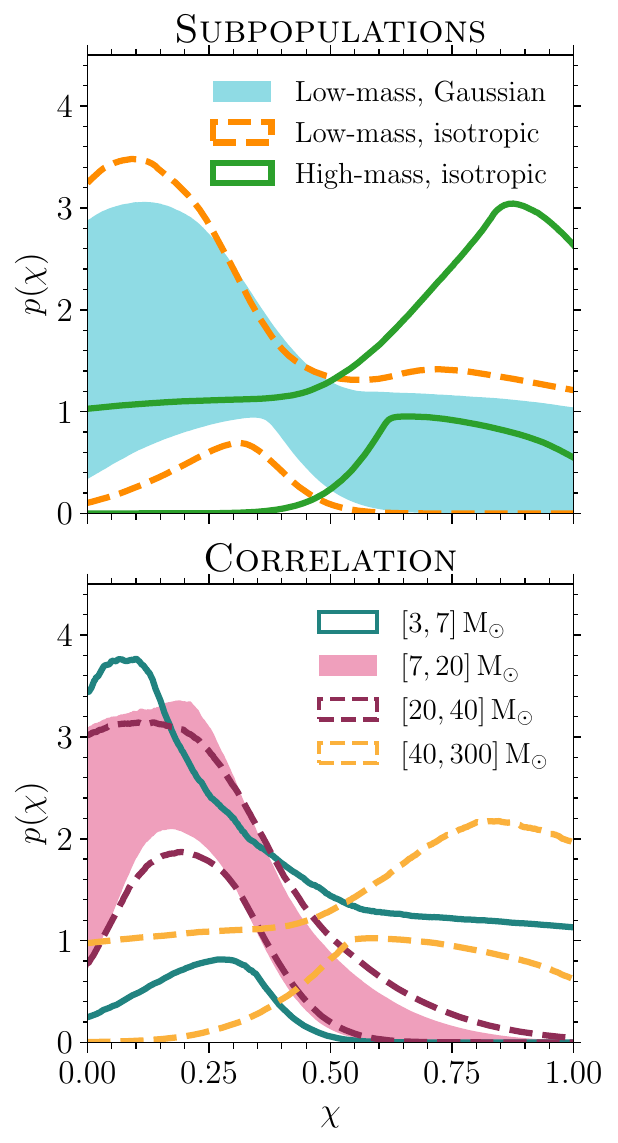}
    \caption{\acp{PPD} on the marginal density of sources as a function of spin magnitude $\chi$ under both models.
    \textit{(Top)} Results obtained with \textsc{Subpopulations} model,
    for \acp{BH} in the low-mass + Gaussian tilts (blue),
    low-mass + isotropic tilts (orange, dashed),
    and high-mass + isotropic tilts (green) subpopulations.
    \textit{(Bottom)} Results obtained under the \textsc{Correlation} model,
    in each bin of \ac{BH} mass.
    Lines or shading enclose the 90\% credible region.
    }
    \label{fig:chi-ppds}
\end{figure}

In the top panel of Fig.~\ref{fig:chi-ppds} we inspect the spin magnitude distributions above and below the transition mass, and associated with different spin tilt distributions under the \textsc{Subpopulations} model.
For \acp{BH} above $\tilde{m}$ ($\sim 40 - 60\,\msun$)
the data are consistent with but do not require a preference for $\chi \sim 0.7$.
Meanwhile, below $\tilde{m}$
the data may prefer small spin magnitudes
for either the isotropic or Gaussian spin tilt subpopulations.

However, under the \textsc{Subpopulations} model,
the spin magnitude distributions for each subpopulation are consistent with one another at 90\% credibility.
We can characterize the spin magnitude distributions
by comparing the density of sources with spin $\chi = 0.1$ vs. $0.7$ over a window $\pm 0.1$,
conditioned on a particular sector $s$ (e.g., subpopulation or mass bin) of the
source parameter space,
\begin{align} \label{eq:Xstat}
    \ln X = \ln \frac{p(0 \leq \chi \leq 0.2 \mid s)}{p(0.6 \leq \chi \leq 0.8 \mid s)} \, .
\end{align}
We find that $\ln X = 2.5^{+3.1}_{-3.1}, 2.5^{+4.9}_{-4.1}, -1.0^{+1.1}_{-9.2}$ for the low-mass + Gaussian, low-mass + isotropic, and high-mass + isotropic subpopulations, respectively.\footnote{We report $\ln X$ instead of $X$ as the fraction of sources becomes small for some ranges of spin magnitudes.}
As these ratios are all consistent with one another,
we cannot conclusively state that any subpopulation prefers large versus small spins.
Broad uncertainties in the spin magnitude distributions
reflect a label swapping degeneracy between both low-mass spin magnitude distributions;
we note that this soft degeneracy arises because spin tilts are not constraining enough
to \textit{solely} identify unique subpopulations.
See App.~\ref{app:sampling} for additional discussion.

Trends between mass and spin magnitude are more certain under the \textsc{Correlation} model.
In the bottom panel of Fig.~\ref{fig:chi-ppds},
we plot posteriors on the marginal density in spin magnitude in each mass bin.
For clarity, we only show the posterior 90\% credible region, but here we also describe which results hold within the 99\% symmetric credible interval.
The spin magnitude distributions between $[7, 20]\,\msun{}$ and $[20, 40]\,\msun{}$ are consistent at 99\% credibility,
while inconsistent with the distribution above $40\,\msun$ at the same credible level.
Meanwhile, the spin magnitude distribution at the lowest masses ($[3, 7]\,\msun{}$) is consistent with all other mass bins at 99\% credibility.
These trends are quantitatively reflected in the fraction of sources
with $\chi \sim 0.1$ vs. $\chi \sim 0.7$ (cf. Eq.~\eqref{eq:Xstat});
we find $\ln X = 0.3^{+7.8}_{-1.1},\,3.7^{+3.4}_{-1.6},\,3.1^{+2.8}_{-1.6},\,$ and $-0.7^{+0.8}_{-3.9}$
for \ac{BH} masses in $[3, 7],\,[7, 20],\,[20, 40]$ and $[40, 300]\,\msun{}$.
Thus, at 90\% credibility, \acp{BH} with mass $\gtrsim 40\,\msun{}$
may prefer large spins (or have equal preference for large vs. small spins)
whereas \acp{BH} with mass between 7 and 40 \msun{}
more often have small spins.
We note that $\ln X$ is poorly measured in the lowest-mass bin;
in the bottom panel of Fig.~\ref{fig:chi-ppds} the 90\% credible region for these masses is similarly broad relative to other bins,
likely reflecting the relatively small number of detected \ac{BBH} with component masses
below $\sim 7\,\msun$.
We show marginal posteriors on the location and scale of the spin magnitude distribution in each mass bin in Fig.~\ref{fig:binned-muchi-sigchi}.

\section{Discussion} \label{sec:discussion}

The masses and spins of merging binary black holes may hint at their formation history,
and we may be able to identify or falsify astrophysical formation channels
by combining catalogs of \ac{BBH} mergers observed via gravitational waves.
Analyses of GWTC-4.0 have identified potential signatures of a subpopulation of relatively heavy ($\gtrsim 45\,\msun$) \acp{BH}
which prefer large, and possibly isotropically-oriented spins,
possibly indicating formation in a dynamical environment \citep{2016ApJ...832L...2R, Gerosa:2021mno, 2022ApJS..258...22R}---although note that some analyses only model the effective binary spin \citep{Antonini:2025ilj, Guttman:2025jkv, Tong:2025xir, Plunkett:2026pxt, Ray:2026uur},
while others explicitly model the component spins \citep{Banagiri:2025dmy, Berti:2025usa, Li:2025iux, Farah:2026jlc, Galaudage:2026opk}.
Simultaneously,
the \acs{LVK}'s analysis of GWTC-4.0 allows for \acp{BH}
to prefer to be oriented in-plane relative to their orbit as opposed to isotropically distributed \citep{LIGOScientific:2025pvj},
which could be indicative of an origin in hierarchical triples (e.g., \citealt{Antonini:2017tgo}).
Recent work by \citealt{Stegmann:2025zkb} suggests that this possible peak in the tilt distribution is associated with relatively light ($\lesssim 45\,\msun$) \acp{BH}.
However, as noted by \citealt{Vitale:2022dpa},
measurements of the population distribution of \ac{BH} spin-orbit (mis)alignments
are sensitive to the assumed form of the population model
(alongside other concerns including numerical and Poisson variance).

In light of these challenges,
we further investigated population-level constraints on structure in the joint
distribution of \ac{BH} masses and spin tilts
with two targeted population models.
In the first model, we augmented the model from \citealt{Stegmann:2025zkb}\footnote{Otherwise, we used the same waveform choices, selection thresholds, and priors as \citealt{Stegmann:2025zkb}.}---which models three distinct subpopulations identified by their mass and spin tilt distributions---allowing for additional flexibility in the marginal primary mass distribution.
We used the same parametric model for primary mass as the \acs{LVK}'s default analysis of GWTC-4.0.
In the second model, we instead allowed for correlations between \ac{BH} masses and spins,
coarsely binning the masses and allowing for different spin magnitude and tilt distributions within each bin.
Using these models,
we made three assessments of \ac{BH} spin orientations, magnitudes, 
and the relationship between \ac{BH} spin and mass with the GWTC-4.0 data:

\textit{There is not
model-independent
evidence for a subpopulation of \acp{BH} with in-plane spins.}
However, we also cannot rule out preference for in-plane spins under either population model (cf. Figs.~\ref{fig:tilts-total-ppd} and \ref{fig:xi}).
When modeling subpopulations in mass and spin,
the data do not require a
subpopulation with a preferred spin orientation.
Neither did we find that in-plane spins were required in the global \ac{BBH} population,
as we inferred a relative fraction of in-plane vs. orbit-aligned spin
tilts between $0.9$ and $3.6$ with 90\% credibility.
Similarly,
at 90\% credibility the data constrain in-plane spins to within a factor of 0.8 to 1.5 of orbit-aligned spins
when we allowed for a correlation between mass and spin.
We found the strongest---but still inconclusive---support for preferentially in-plane spins for \acp{BH} (primary or secondary) with masses between $20$ and $40\,\msun$ (cf. Fig.~\ref{fig:tau-per-bin}).

\textit{There is not a statistically-significant correlation between \ac{BH} masses and spin orientation.}
When modeling subpopulations in mass and spin,
we inferred a transition in the spin distribution at $\sim 45\,\msun{}$;
we assumed an isotropic distribution of tilts above this threshold,
and below we inferred a spin tilt distribution consistent with isotropy
(cf. Fig.~\ref{fig:xi}).
When modeling a correlation between mass and spin,
we found that the spin tilt distributions
at all mass scales were consistent with isotropy at 90\% credibility
(cf. Fig.~\ref{fig:tau-per-bin}).
While there may be hints that black holes between $7$ and $20\,\msun$
prefer orbit-aligned spin-tilts---consistent with typical models of field formation---we cannot rule out that spins tend to be randomly oriented at all masses,
corroborating similar findings by \citealt{ Banagiri:2025dmy}, \citealt{Berti:2025usa} and \citealt{Farah:2026jlc}.
Given that our results are consistent with both isotropy and peaks in the spin tilt distribution,
in App.~\ref{app:inj} we show that finite catalogs drawn from isotropic spin populations can spuriously yield a preference for in-plane spin tilts, complementing \citealt{Vitale:2025lms} which demonstrated that spurious features in the tilt distribution can also arise from preferentially orbit-aligned populations.

\textit{There is a statistically-significant correlation between \ac{BH} masses and spin magnitudes.}
When we allowed a correlation between \ac{BH} mass and spin,
we found that \acp{BH} above $40\,\msun{}$ \textit{may} tend to have larger ($\gtrsim 0.6$)
spin magnitudes and \acp{BH} below $40\,\msun{}$ \textit{confidently} tend
to have smaller ($\lesssim 0.2$) spin magnitudes (bottom panel of Fig.~\ref{fig:chi-ppds}).
This corroborates findings by \citealt{Banagiri:2025dmy},  \citealt{Berti:2025usa}, and \citealt{Farah:2026jlc}.
Notably, however, we did \textit{not} find significant evidence for distinct spin magnitude
distributions at different masses when modeling subpopulations in primary mass--spin tilt (top panel of Fig.~\ref{fig:chi-ppds}).

Our work adds a cautionary tale to a growing body of literature regarding
population-level constraints on \ac{BH} spin orientations.
Despite a more-than-doubling of the catalog of \acp{BBH} since the end of
the previous \ac{LVK} observing run,
the observed data are still insufficient to draw (or enforce)
astrophysical conclusions based on spin tilts \textit{alone}.
As demonstrated here, spin tilt population inference
is intimately related to population-level assumptions
in other sectors of the source parameter space.
Given poor single-event measurements on the spin tilts,
it may be that population-level measurements are ultimately driven
by constraints on the spin tilts provided by some combination of the
effective spin, component spin magnitudes, and masses.
If so, robust inference of \ac{BH} spin orientations
would require population models that can respect the constraints
imposed by other, better-measured source parameters.
In turn,
to search for subpopulations with different spin tilt distributions,
each subpopulation should be modeled with its own spin magnitude \textit{and} mass distributions.
Yet, we also found that models that attempt to distinguish populations primarily from their spin properties can present significant challenges during sampling (cf. App.~\ref{app:sampling}).
Instead,
models that allow for correlations in a hierarchy from better-measured to poorly-measured parameters
like our model which independently correlates spin magnitude and tilt with mass (cf. Sec.~\ref{sec:corr}),
may enable more robust population inference.

While we focused on only two model variations,
one of our key conclusions---that spin tilts are essentially unconstrained at all masses---is relevant when comparing population models.
Population models are often compared by their Bayes' factors,
which include the prior volumes associated with each model.
That prior volume is reduced if we fix the spin tilt distribution for all \acp{BH} or a particular subpopulation,
increasing the Bayes' factor relative to an analysis which makes no such assumptions about tilts.

The lack of confident mass--tilt subpopulations or correlations is also relevant when astrophysically interpreting analyses that model the effective spins of binaries.
While multiple works have identified transitions in the effective spin distribution(s) as a function of primary mass \citep{Antonini:2025ilj, Guttman:2025jkv, Tong:2025xir, Plunkett:2026pxt, Ray:2026uur},
different mass and component spin distributions can yield the same effective spin distribution \citep{Miller:2024sui}.
Instead, the component masses and spins are easier to interpret in the context of the formation history of each \ac{BH} in a binary.
Our results corroborate a preference for lighter \acp{BH} to have smaller spins
which could be indicative of field formation,
or alternatively first-generation mergers in dynamical environments  \citep{Rodriguez:2019huv},
although these interpretations rely on other uncertain astrophysics like angular momentum transport in massive stars \citep{Fuller:2019sxi}
or the role of accretion in the mass and spin evolution of \acp{BH}, e.g., \citealt{Kiroglu:2025vqy, Olejak:2026nhy}.
Identifying those same \acp{BH} as nearly always having at most a few degrees of spin-orbit misalignment would lend additional credence to the identification of
an isolated binary subpopulation (although see \citealt{Baibhav:2024rkn} for an overview of theoretical pathways to forming misaligned spins in field binaries).
Ultimately,
careful population modeling,
triple checking the solidity of results against analysis choices,
and more data will be required to confidently distinguish black hole formation channels from their spin tilts.

\begin{acknowledgments}

We are grateful to Jakob Stegmann for insightful discussions about his work and making his analysis publicly available.
We also thank Sof\'ia \'Alvarez-L\'opez, Sharan Banagiri, Paul Draghis, Carl-Johan Haster, Jack Heinzel, Asad Hussain, Utkarsh Mali, Matthew Mould, Cailin Plunkett for discussions and Sylvia Biscoveanu for internal review.
We thank the anonymous referee for a thoughtful and constructive review.
N.E.W. is supported by the National Science Foundation Graduate Research Fellowship Program under grant No. 2141064.
S.V. is partially supported by the NSF grant No. PHY-2045740.
M.Z. gratefully acknowledges funding from the Brinson Foundation in support of astrophysics research at the Adler Planetarium. 
The authors are grateful for computational resources provided by
the LIGO Laboratory supported by National Science Foundation Grants PHY-0757058 and PHY-0823459,
and by subMIT at MIT Physics.
This material is based upon work supported by NSF's LIGO Laboratory which is a major facility fully funded by the National Science Foundation and has made use of data or software obtained from the Gravitational Wave Open Science Center (gwosc.org), a service of the LIGO Scientific Collaboration, the Virgo Collaboration, and KAGRA.

\end{acknowledgments}

\appendix

\section{Models and priors} \label{app:models}

Here, we describe the functional form of the \textsc{Subpopulations} and \textsc{Correlation} population models.
We use a subscript ``1" and ``2" to denote parameters of the primary and secondary \acp{BH} in each binary,
with masses $m_1$ and $m_2$ and ratio $q = m_2 /m_1 \leq 1$.
For brevity, throughout we suppress the parameters of the population models unless stated explicitly.

\subsection{\textsc{Subpopulations} model}

Under the \textsc{Subpopulations} model, following \citealt{Stegmann:2025zkb} we model the spin magnitudes and tilts as a three-component mixture,
\begin{equation} \label{eq:subpop-model}
\begin{split}
    p(\chi_1, \chi_2, \cos \tau_1, \cos \tau_2 \mid m_1) &= \bigg[ \zeta(m_1)\,p(\chi_1, \chi_2, \cos \tau_1, \cos \tau_2 \mid \text{high-mass, isotropic}) \\
    &+ (1-\zeta(m_1))\xi\,p(\chi_1, \chi_2, \cos \tau_1, \cos \tau_2 \mid \text{low-mass, Gaussian}) \\
    &+ (1-\zeta(m_1))(1- \xi)\,p(\chi_1, \chi_2, \cos \tau_1, \cos \tau_2 \mid \text{low-mass, isotropic}) \bigg] \, , \\
\end{split}
\end{equation}
where $\zeta$ is a sigmoid function of $m_1$ such that $\zeta$ is large when $m_1 > \tilde{m}$ and small otherwise,
\begin{equation}
    \zeta(m_1) = \left[ 1 + \exp(\tilde{m} - m_1) \right]^{-1} \, .
\end{equation}
We label each subpopulation as ``low-mass'' or ``high-mass'' depending on whether it models \acp{BBH} with primary mass falling below or above $\tilde{m}$.
We also label subpopulations based on how cosine-spin tilts are distributed: ``isotropic'' indicating a uniform distribution and ``Gaussian'' for a truncated Gaussian distribution.
Within each subpopulation, spin magnitudes are modeled as truncated Gaussians;
subpopulations do \textit{not} share the same spin magnitude distributions.
The primary and secondary spin magnitudes and tilts are also identically distributed within each subpopulation.
We can re-write Eqn.~\eqref{eq:subpop-model} with these modeling choices;
for clarity, we also include the parameters of each spin distribution,
\begin{equation} \label{eq:subpop-model-full}
\begin{split}
    p(\chi_1, \chi_2, &\cos \tau_1, \cos \tau_2 \mid m_1) = \bigg[ \zeta(m_1)\,\frac{1}{4}\mathcal{N}_{[0, 1]}(\chi_1 \mid \mu_\chi^\text{High Iso}, \sigma_\chi^\text{High Iso}) \mathcal{N}_{[0, 1]}(\chi_2 \mid \mu_\chi^\text{High Iso}, \sigma_\chi^\text{High Iso}) \\
    &+ (1-\zeta(m_1))\xi\,\mathcal{N}_{[-1, 1]}(\cos \tau_1 \mid \mu_\tau, \sigma_\tau) \mathcal{N}_{[-1, 1]}(\cos \tau_2 \mid \mu_\tau, \sigma_\tau) \mathcal{N}_{[0, 1]}(\chi_1 \mid \mu_\chi, \sigma_\chi) \mathcal{N}_{[0, 1]}(\chi_2 \mid \mu_\chi, \sigma_\chi) \\
    &+ (1-\zeta(m_1))(1- \xi)\,\frac{1}{4} \mathcal{N}_{[0, 1]}(\chi_1 \mid \mu_\chi^{\mathrm{Iso}}, \sigma_\chi^\mathrm{Iso}) \mathcal{N}_{[0, 1]}(\chi_2 \mid \mu_\chi^{\mathrm{Iso}}, \sigma_\chi^\mathrm{Iso}) \bigg]\ \, , \\
\end{split}
\end{equation}
where $\mathcal{N}_{[a,b]}(\cdot \mid \mu, \sigma)$ denotes a truncated Gaussian distribution on a domain $[a, b]$ with location $\mu$ and scale $\sigma$.
Parameters with superscript ``High Iso'' or ``Iso'' control the shape of the ``high-mass, isotropic'' or ``low-mass, isotropic'' subpopulations, respectively.
Parameters without a superscript control the low-mass, Gaussian subpopulation.
The entire population model can be written,
\begin{equation}
    p(z)p(m_1, q)p(\chi_1, \chi_2, \cos \tau_1, \cos \tau_2 \mid m_1)
\end{equation}
where $p(z)$ is derived from a power law in the differential merger rate density \citep{Fishbach:2018edt} and $p(m_1, q)$ is the \textsc{Broken Power Law Plus Two Peaks} from \citet{LIGOScientific:2025pvj}.

\subsection{\textsc{Correlation} model}

Under the \textsc{Correlation} model, the spin magnitudes and tilts are written
\begin{align}
    p(\chi_1, \chi_2 \mid m_1, q) &= p(\chi_1 \mid m_1) p(\chi_2 \mid m_2) \, , \\
     p(\cos \tau_1, \cos \tau_2 \mid m_1, q) &= p(\cos \tau_1 \mid m_1) p(\cos \tau_2 \mid m_2) \, ,
\end{align}
where $m_2 = q\,m_1$ and the spin magnitude and tilt distributions follow Eqns.~\eqref{eq:correlation-spin-mag} and \eqref{eq:correlation-spin-tilt}.
Thus, the entire population model can be written
\begin{align}
    p(z) p(m_1, q) p(\chi_1, \chi_2 \mid m_1, q) p(\cos \tau_1, \cos \tau_2 \mid m_1, q) \, ,
\end{align}
where $p(z)$, $p(m_1, q)$ have the same functional form as in the \textsc{Subpopulations} model.

\subsection{Priors}

Priors on the parameters of spin magnitude and tilt under each model are listed in Tab.~\ref{tab:spin-priors}.
Priors on the parameters of the mass and redshift distributions are listed in Tab.~\ref{tab:mass-priors}.
For the \textsc{Subpopulations} model,
we also adopt a uniform prior on the transition mass $\tilde{m}$ over $[10, 100]\,\msun$.

\begin{table}[h!]
    \centering
    \begin{tabular}{l c c}
        \toprule
        \toprule
        \textbf{Parameter} & \textbf{Symbol} & \textbf{Prior} \\
        \midrule
        Location & $\mu_\chi$ & $[0, 1]$ \\
        Scale & $\sigma_\chi$ & $[0.1, 1]$ \\
        \midrule
        \textsc{Subpopulations} \\
        Peak location & $\mu_\tau$ & $[-1, 1]$ \\
        Peak width & $\sigma_\tau$ & $\mathcal{N}_{[0.1, 4]}$ \\
        Peak mixing fraction & $\xi$ & $[0, 1]$ \\ \\
        \textsc{Correlation} & & \\
        Peak location & $\mu_\tau$ & $[-1, 1]$ \\
        Peak width & $\sigma_\tau$ & $[0.01, 4]$ \\
        Peak mixing fraction & $\xi$ & $[0, 1]$ \\
        \bottomrule
        \bottomrule
    \end{tabular}
    \caption{
    Priors on the parameters specifying the astrophysical distribution of
    spin magnitudes (top) and tilts (bottom).
    Under the \textsc{Subpopulations} model, each subpopulation has distinct spin magnitude parameters inferred with the listed priors.
    Above the transition mass, the spin tilt distribution is assumed isotropic;
    below, it is a mixture between isotropic and Gaussian with mixing fraction $\xi$.
    Here, $\mathcal{N}_{[0.1, 4]}$ denotes a truncated Gaussian prior with location $\mu = 0$ and scale $\sigma = 1/2$ following \citet{Stegmann:2025zkb}.
    Under the \textsc{Correlation} model,
    each mass bin has a distinct spin magnitude and spin tilt distribution with the listed priors;
    the spin tilt distribution in each mass bin is also modeled as a mixture between isotropic and Gaussian.
    Note that the same spin distribution parameters are used to model all \acp{BH} falling in a given mass bin,
    and so there are not distinct spin parameters for primary vs. secondary \ac{BH} spins.
    We denote a uniform prior between $a$ and $b$ as $[a, b]$.
    }
    \label{tab:spin-priors}
\end{table}
\begin{table}[h!]
    \centering
    \begin{tabular}{ l c c }
        \toprule
        \toprule
        \textbf{Parameter} & \textbf{Symbol} & \textbf{Prior} \\
        \midrule
        $m_1$ low-mass power law index & $\alpha_1$ & $[-4, 12]$ \\
        $m_1$ high-mass power law index & $\alpha_2$ & $[-4, 12]$ \\
        Break mass & $m_{\rm break}$ & $[20, 50]\,M_\odot$ \\
        First peak location & $\mu_{1}$ & $[5, 20]\,M_\odot$ \\
        First peak width & $\sigma_{1}$ & $[0, 10]\,M_\odot$ \\
        Second peak location & $\mu_{2}$ & $[25, 60]\,M_\odot$ \\
        Second peak width & $\sigma_{2}$ & $[0, 10]\,M_\odot$ \\
        Low-mass smoothing scale & $\delta_m$ & $[0, 10]\,M_\odot$ \\
        Minimum \ac{BH} mass & $m_{\min{}}$ & $[3, 6.5]\,M_\odot$ \\
        Power law, peak mixing fractions & $\lambda_0, \lambda_1$ & Dirichlet \\
        $q$ power law index & $\beta$ & $[-2, 7]$ \\
        \midrule
        $z$ power law index & $\lambda_z$ & $[-6, 6]$ \\
        \bottomrule
        \bottomrule
    \end{tabular}
    \caption{
    Priors on the parameters specifying the astrophysical distribution of
    masses (top) and redshifts (bottom).
    We denote a uniform prior between $a$ and $b$ as $[a, b]$.
    The Dirichlet prior is equally-weighted between all components.
    }
    \label{tab:mass-priors}
\end{table}

\begin{figure*}
    \centering
    \includegraphics[width=0.7\linewidth]{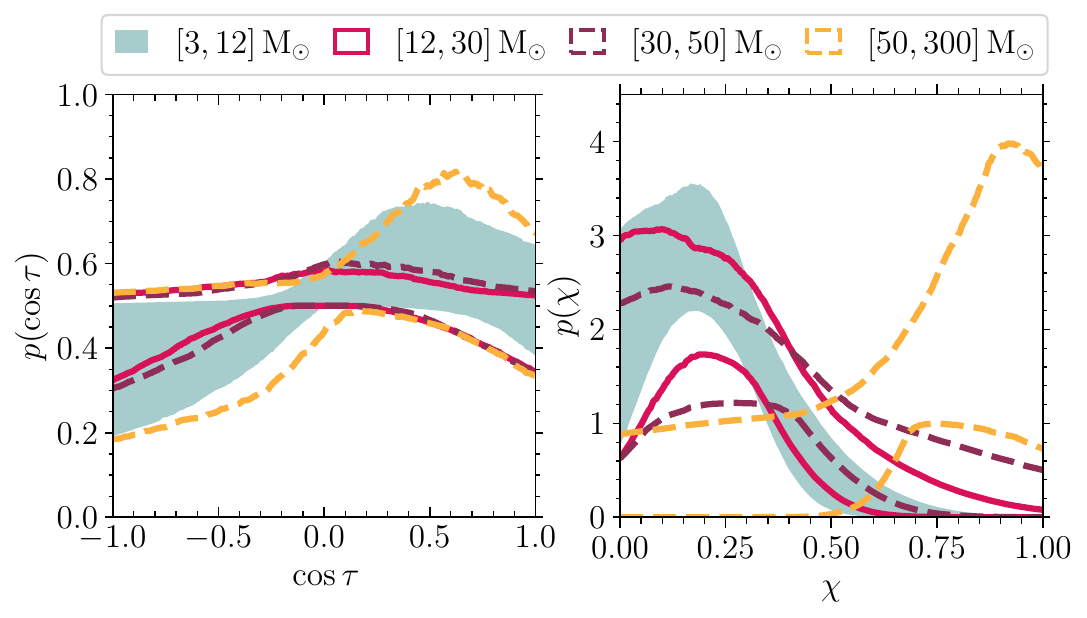}
    \caption{\acp{PPD} on the marginal density of sources as a function of spin tilt $\cos \tau$ (left) and magnitude $\chi$ (right) in four bins of \ac{BH} mass under the \textsc{Correlation} model, using a different choice of bin edges than the results presented in Sec.~\ref{sec:results}.
    }
    \label{fig:altbins}
\end{figure*}

\begin{figure*}
    \centering
    \includegraphics[width=0.98\linewidth]{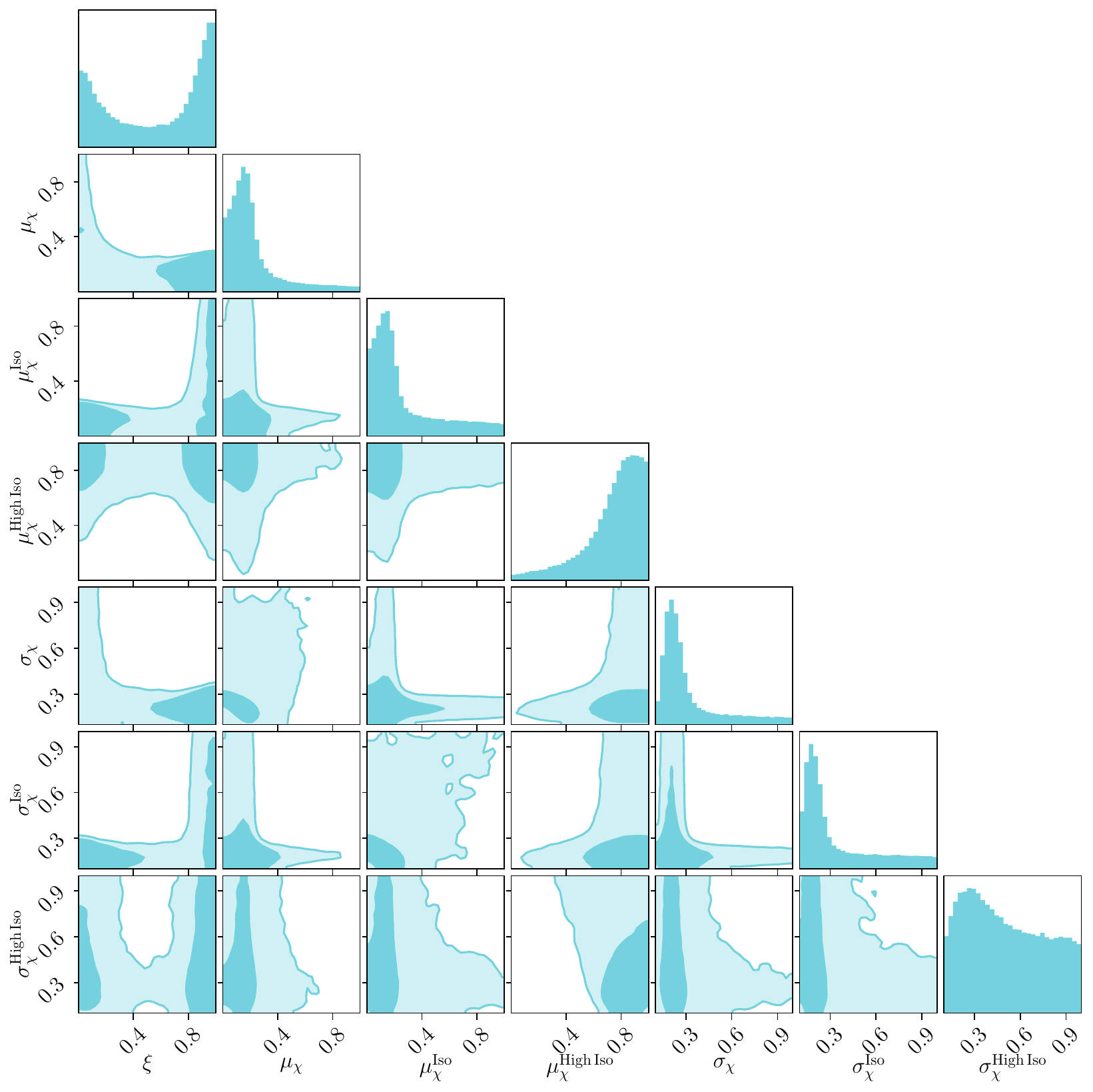}
    \caption{Joint marginal posterior on the branching ratio $\xi$ between low-mass isotropic and Gaussian spin tilt distributions under the \textsc{Subpopulations} model,
    as well as the location ($\mu_\chi$, $\mu_\chi^\text{Iso}$, $\sigma_\chi^\text{High Iso}$) and scale ($\sigma_\chi$, $\sigma_\chi^\text{Iso}$, $\sigma_\chi^\text{High Iso}$) parameters of the spin distributions for each respective subpopulation.
    Contours shown enclose the 50\% and 90\% credible regions.
    }
    \label{fig:subpop-xi-muchi-sigchi}
\end{figure*}

\begin{figure*}
    \centering
    \includegraphics[width=0.98\linewidth]{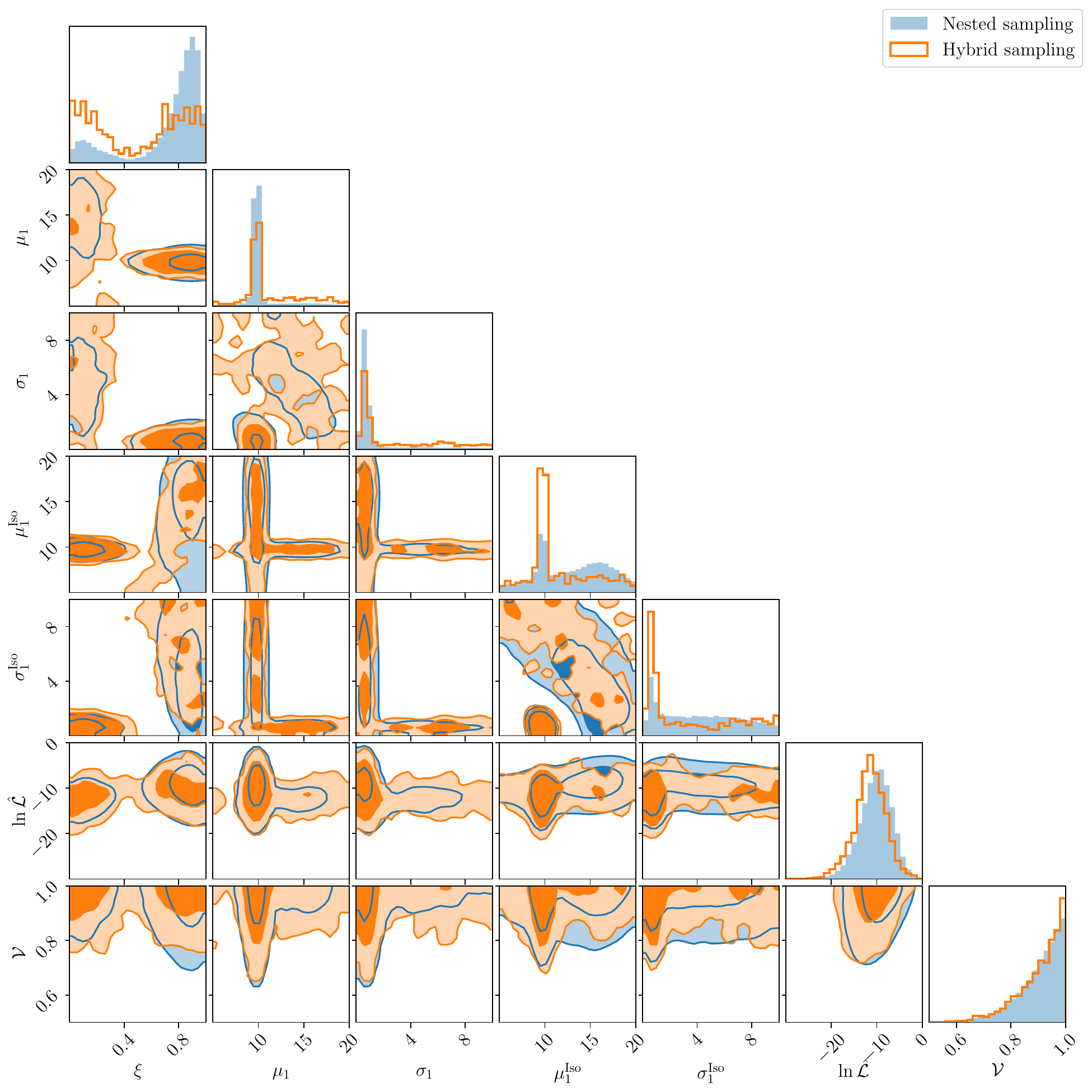}
    \caption{
    Marginal posteriors on the branching ratio $\xi$ between truncated Gaussian and isotropic spin tilt distributions (below the transition mass),
    as well as the location ($\mu_1, \mu_1^\text{Iso}$) and scale ($\sigma_1, \sigma_1^\text{Iso}$) of the first peak in the primary mass distributions,
    under a variation of the \textsc{Subpopulations} model where the truncated Gaussian vs. isotropic tilt distributions each have their own mass distribution (cf. App.~\ref{app:multimass}).
    We also show the log-likelihood estimator $\ln \hat{\mathcal{L}}$ and the variance $\mathcal{V}$ of that estimator.
    Log-likelihoods are shown relative to the maximum likelihood.
    }
    \label{fig:threemass}
\end{figure*}

\section{Alternative mass bins} \label{app:altbins}

Here, we test whether a variation in the mass bin edges in the \textsc{Correlation} model qualitatively changes the result described in Sec.~\ref{sec:results}.
Our original binning choice includes two bins, spanning $[7, 20]\,\msun$ and $[20, 40]\,\msun$,
chosen to straddle the $\sim10\,\msun$ and $\sim35\,\msun$ features in the mass distribution.
However, there may also be a power law continuum between these features extending from roughly $12$ to $30$ \msun \citep{LIGOScientific:2025pvj},
and \acp{BH} that might be associated with that continuum rather than the peaks are captured in the $[7, 20]\,\msun$ and $[20, 40]\,\msun$ bins.
We repeat our analysis with bin edges chosen to provide one bin each for the $\sim 10\,\msun$ peak, $\sim35\,\msun$ feature, and power law continuum, taking bins of $[3, 12]$, $[12, 30]$, $[30, 50]$, and $[50, 300]\,\msun$.

In Fig.~\ref{fig:altbins} we show 90\% credible regions of \acp{PPD} of spin tilts and magnitudes within each mass bin.
Looking first at tilts,
in the left panel
we find that within the $[3, 12]\,\msun$ and $[50, 300]\,\msun$ bins there is support for broad peaks away from in-plane spins,
while in between---from $12$ to $50$\,\msun{}---only slight deviations from isotropy are allowed within the  90\% credible region.
Qualitatively, these results agree with the \acp{PPD} under our original choice of bins (cf. Fig.~\ref{fig:tau-per-bin});
under both choices, the bin including the $10\,\msun$ peak and the highest mass bin both have support for some degree of orbit-alignment.
However, we note that under our original bin choices the $[20, 40]\,\msun$ bin had support for a peak near $\cos \tau \sim 0$ which is not found in either the $[12, 30]\,\msun$ or $[30, 50]\,\msun$ bins.

Considering spin magnitudes,
in the right panel of Fig.~\ref{fig:altbins}
we find that \acp{BH} below $50\,\msun$ tend to have small spins ($\chi \lesssim 0.5$)
whereas \acp{BH} above $50\,\msun$ may tend to have large spins.
Qualitatively, the conclusion that heavier black holes may tend to spin faster than lighter black holes agrees with the bottom panel of Fig.~\ref{fig:chi-ppds}.
In detail, Fig.~\ref{fig:altbins} also shows that the spin magnitude distribution may broaden as the mass increases from the $[3, 12]\,\msun$ bin through to the $[30, 50]\,\msun$ bin,
whereas in Fig.~\ref{fig:chi-ppds} \acp{BH} from $7$--$40\,\msun$ have the same spin magnitude distribution within 90\% credibility.

Overall, our key conclusions remain unchanged under a variation in bin edges:
there is not a statistically-significant peak at $\cos \tau \sim 0$
and there is not a statistically-significant correlation between spin tilts and mass,
while there does appear to be correlation between mass and spin magnitude where heavier black holes tend to spin faster.

\section{Sampling and label degeneracy} \label{app:sampling}

\subsection{Nested sampling settings}

To produce the results in this work, we used $10^4$ live points under the \textsc{Subpopulations} model and $10^3$ under the \textsc{Correlation} model.
Likelihood iso-contours are sampled according to the \texttt{acceptance-walk} method
with \texttt{naccept}=5 (argument names are as implemented in \texttt{bilby}; see \citealt{bilby_paper}).

\subsection{Degenerate subpopulations}

While testing the \textsc{Subpopulations} model,
we noticed that changing the random seed provided to \texttt{dynesty} yielded different marginal posteriors on the branching ratio $\xi$ between the low-mass isotropic and Gaussian spin tilt distributions.
In roughly half of analyses with different seeds,
we recovered a single mode at $\xi \sim 1$,
and in the other half, at $\xi \sim 0$;
this was despite using 1000 live points.
Combining the result of 10 nested sampling analyses
with 1000 live points each (evidence-weighting the posterior samples to create a combined sample set),
we recovered both modes with similar likelihoods.
We verified this result with an additional run with $10^4$ live points;
that same analysis is presented in the main body for results under the \textsc{Subpopulations} model.

After further inspection of the population posterior,
we identified that each mode in $\xi$ was associated with a degenerate label-swap between the locations $\mu_\chi$, $\mu_\chi^\text{Iso}$
and scales $\sigma_\chi, \sigma_\chi^\text{Iso}$
of the spin distributions associated with the low-mass Gaussian and low-mass isotropic tilt distributions,
respectively.
The joint marginal on $(\xi, \mu_\chi, \mu_\chi^\text{Iso}, \sigma_\chi, \sigma_\chi^\text{Iso})$ is shown in Fig.~\ref{fig:subpop-xi-muchi-sigchi} (where we also include the location $\mu_\chi^\text{High Iso}$ and scale $\sigma_\chi^\text{High Iso}$ of the spin magnitude distribution in high-mass, isotropic-tilt subpopulation).

\subsection{Multiple mass models} \label{app:multimass}

We also experimented with variations on the \textsc{Subpopulations} model,
where each tilt--identified subpopulation was allowed to have its own mass distribution (i.e. \textsc{Broken Power law plus Peak} functional form but parameters are not shared between subpopulations).
We use the same spin priors as the \textsc{Subpopulations} model (see App.~\ref{app:models}), most notably a truncated normal prior on the scale of the truncated Gaussian in spin tilts.
To validate our results,
we inferred the population parameters using nested sampling as well as a hybrid variant of parallel-tempered MCMC (``hybrid sampling''; \citealt{Wolfe:2022nkv}) based on \texttt{ptemcee} \citep{2013PASP..125..306F, 2016MNRAS.455.1919V}. 
Using nested sampling,
we performed 10 analyses initialized with different random seeds and jointly resampled these results, weighted by their evidence.
We initialized hybrid sampling from one of these nested sampling runs which had support for both $\xi \sim 0$ and $1$;
we used 200 walkers and 10 temperatures.

In Fig.~\ref{fig:threemass} we plot marginal posteriors
on $\xi$ jointly with the location ($\mu_1$, $\mu_1^\text{Iso}$) and scale ($\sigma_1$, $\sigma_1^\text{Iso}$) of the first peak in the mass distribution associated with each subpopulation,
when allowing each subpopulation to have its own mass distribution.
We also show the log-likelihood estimator $\ln \hat{\mathcal{L}}$
and the variance $\mathcal{V}$ of that estimator.
We found one mode where the low-mass, truncated Gaussian spin tilt subpopulation has a peak in the primary mass distribution at $10\,\msun$
($\mu_1 \sim 10\,\msun$, $\sigma_1 \lesssim 1\,\msun$),
and another mode where that peak is associated with the low-mass, isotropic spin tilt subpopulation
($\mu_1^\text{Iso} \sim 10\,\msun$, $\sigma_1^\text{Iso} \lesssim 10\,\msun$).
These modes correspond to $\xi \sim 1$ and $\xi \sim 0$, respectively.
This multi-modality appears as a label degeneracy
in the joint marginals of $(\mu_1, \mu_1^\text{Iso})$ and $(\sigma_1, \sigma_1^\text{Iso})$,
creating ridges of roughly equal log-likelihood in the population parameter space.
These geometries (multi-modality and ridges) are challenging to sample;
the two sampling algorithms we employed did not agree on the final result,
particularly in the tails of the posterior,
even though both algorithms terminated according to their own internal convergence criteria.
Inference under label degenerate models is an active area of research; e.g., \citet{2019PhRvD.100h4041B}.

\section{Simulated isotropic population} \label{app:inj}

\begin{figure}
    \centering
    \includegraphics[width=0.49\linewidth]{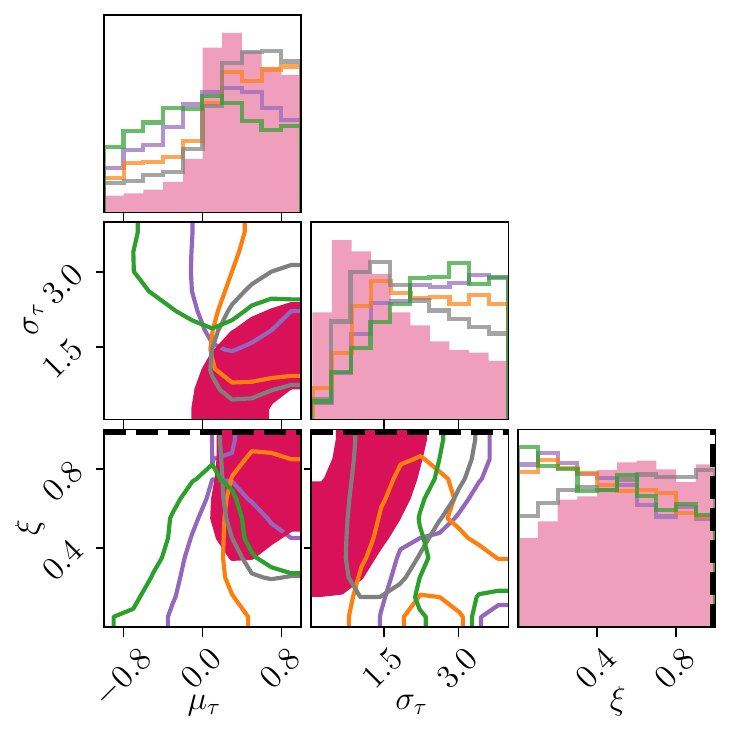} \hfill
    \includegraphics[width=0.49\linewidth]{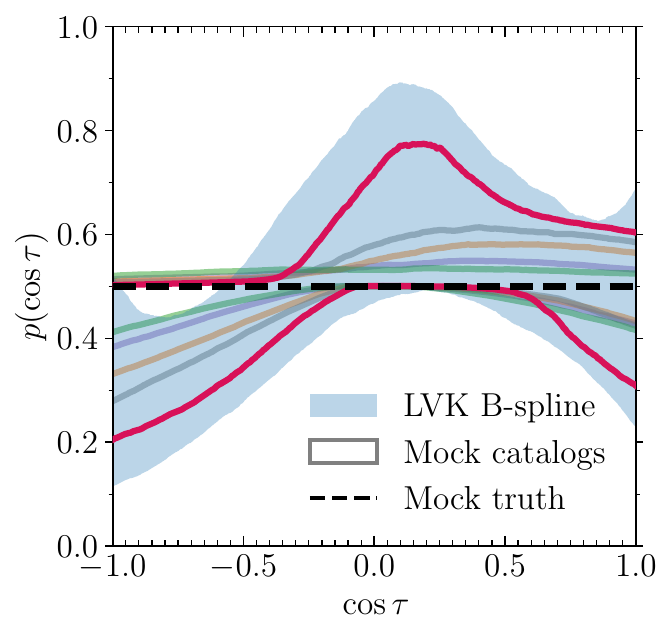}
    \caption{
    (\textit{Left}) Marginal posteriors on the location, scale, and mixing fraction (relative to an isotropic component) of the truncated Gaussian distribution in tilts given five different mock catalogs.
    Lines in the joint marginals only enclose the 50\% credible region for clarity.
    (\textit{Right}) \ac{PPD} in spin tilt for the same analyses.
    The true astrophysical distribution is isotropic (black, dashed).
    We highlight the mock catalog which returns the strongest support for a peak in red;
    analyses of four other mock catalogs are shown with other colors (green, orange, purple, gray).
    For comparison, in blue we include the \acs{LVK}'s analysis using B-splines \citep{Edelman:2022ydv, LIGOScientific:2025pvj}.
    }
    \label{fig:inj-tilts}
\end{figure}

\begin{figure}
    \centering
    \includegraphics[width=0.35\linewidth]{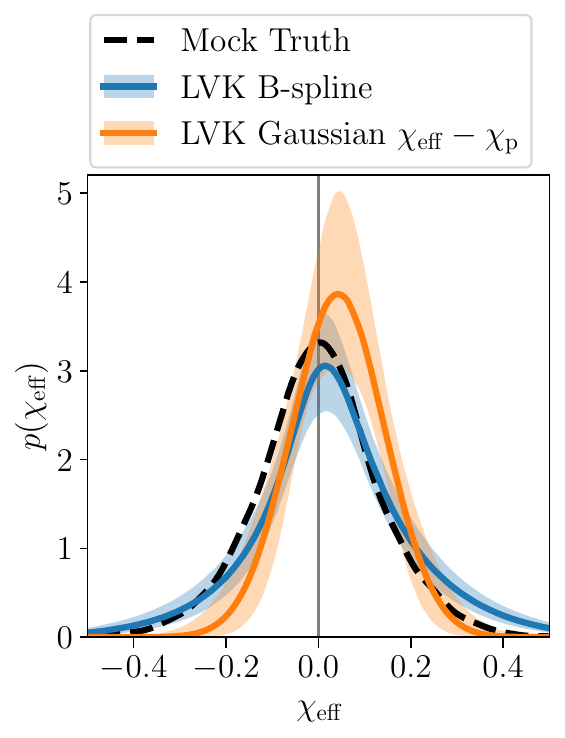}
    \caption{
    Simulated effective spin $\chi_\mathrm{eff}$ distribution used for mock catalog realizations (black)
    compared to effective spin \acp{PPD} recovered by two \ac{LVK} analyses:
    the \ac{PPD} implied by the B-spline analysis (blue),
    and the marginal \ac{PPD} when directly modeling the population distribution of $\chi_\mathrm{eff}$ and effective precessing spin $\chi_\mathrm{p}$ as a bivariate Gaussian (orange) \citep{LIGOScientific:2025pvj}.
    Medians are denoted with solid lines and shaded regions enclose the 90\% credible region.
    }
    \label{fig:chieff}
\end{figure}

\citealt{Vitale:2025lms} showed that orbit-aligned spin populations can spuriously imply isotropic tilts or peaks away from $\cos \tau \sim 1$.
Here we test the converse: whether spins drawn from an isotropic distribution can yield spurious peaks.
We simulate 500 merging \ac{BBH}
with masses, redshifts, and spin magnitudes drawn according to the preferred model from the \acs{LVK}'s analysis of GWTC-3 \citep{KAGRA:2021duu};
meanwhile, spin tilts are isotropically-distributed.
See \citealt{Vitale:2025lms} and \citealt{Wolfe:2025yxu} for additional details on the
simulated population, parameter estimation, and sensitivity estimation.

We perform inference on five mock catalogs of 150 sources each,
drawn without repetition from the total set of 500 simulations;
note that these catalogs are not independent and sometimes contain the same events.
During inference,
we adopt the same model as the astrophysical distribution:
a smoothed power-law plus Gaussian peak in primary mass;
a power law in mass ratio;
a power law in the differential merger rate density as a function of redshift;
and a beta distribution in spin magnitudes (identical for both components).
For spin tilts, we adopt a mixture of isotropic and Gaussian truncated distributions in spin tilt, also identical for both components.
We followed the priors in Tab.~III of \citealt{Wolfe:2025yxu},
except we set uniform priors over $[1, 10]$ directly on the alpha and beta parameters of the spin magnitude distribution,
rather than its mean and variance.
We also take a uniform prior over $[0.01, 4]$ for $\sigma_\tau$ to match all other analyses in this work.
We sample the population posterior with \texttt{dynesty} as in the rest of this work,
using 500 live points.

In the left panel of Fig.~\ref{fig:inj-tilts} we show marginal posteriors on $\mu_\tau$, $\sigma_\tau$, and $\xi$ (parameter symbols follow Tab.~\ref{tab:spin-priors}).
In the right panel,
we show \acp{PPD} in cosine-spin tilt for all mock catalog realizations,
as well as the \acs{LVK}'s analysis of GWTC-4.0 \citep{LIGOScientific:2025pvj, ligo_scientific_collaboration_2025_16911563} using B-splines \citep{Edelman:2022ydv}.
In red, we highlight the result with the strongest support for a relatively narrow peak in spin tilts near $\cos \tau \sim 0$.
The 90\% credible bounds inferred for this particular mock catalog appear similar to the bounds found for GWTC-4.0 under the B-spline model.
In detail,
the upper bound for our highlighted mock analysis peaks with a marginal density of $p(\cos \tau) \sim 0.8$ at $\cos \tau \sim 0.2$,
and the upper bound for the LVK B-spline analysis peaks with $p(\cos \tau) \sim 0.9$ at $\cos \tau \sim 0.1$.
We note that the B-spline model is more flexible than an isotropic plus Gaussian mixture;
in turn, the posterior uncertainty on the B-spline model will tend to be larger.
That we can recover spurious support for a peak near $\cos \tau \sim 0$---even under a relatively strong model---further indicates that GWTC-4.0 does not yet demand a preferred spin orientation.

We note that an isotropic tilt distribution is inconsistent with constraints on the effective spins from GWTC-4.0.
Larger effective spins are typically easier to measure \citep{Vitale:2016avz, Chatziioannou:2018wqx, Shaik:2019dym}.
If constraints on spin tilts are typically driven by constraints on effective spins,
our mock catalogs may be optimistic (pessimistic) about the measurability of tilts at the population level
and less (more) sensitive to Poisson catalog variations
if our simulated population contains more (fewer) sources with large $\chi_\mathrm{eff}$ than reality.
In Fig.~\ref{fig:chieff} we compare the distribution of effective spins in our simulated population to the effective spin distribution implied by the LVK GWTC-4.0 B-spline analysis (which directly models the component spin and mass ratio distributions),
as well as an \ac{LVK} GWTC-4.0 analysis which jointly models $\chi_\mathrm{eff}$ and the effective precessing spin $\chi_\mathrm{p}$ as a bivariate Gaussian \citep{LIGOScientific:2025pvj}.
We find that our simulated population tends to have fewer sources with $0 \lesssim \chi_\mathrm{eff} \lesssim 0.2$ compared to the distribution recovered with the Gaussian effective spin model,
although both agree for $\chi_\mathrm{eff} \gtrsim 0.2$.
Simultaneously, our simulated population is consistent with the B-spline analysis for $0 \lesssim \chi_\mathrm{eff} \lesssim 0.2$,
and the bulk of the simulated population lies within or near the 90\% credible region for the B-spline analysis.

\section{Additional plots}

\begin{figure}
    \centering
    \includegraphics[width=0.6\linewidth]{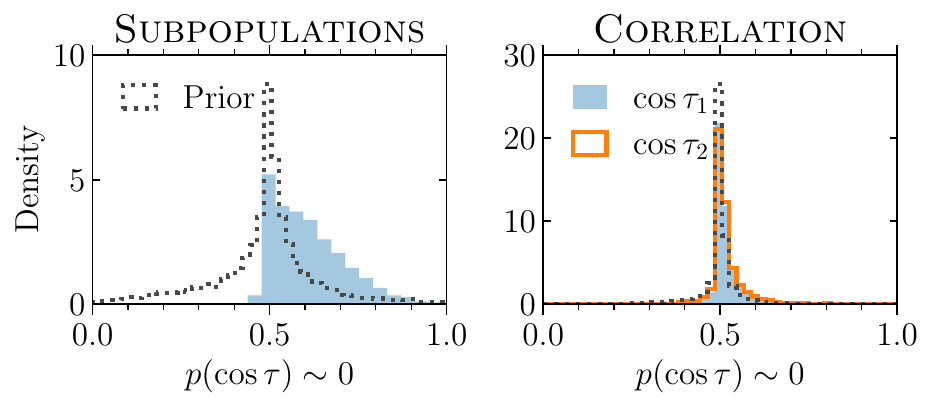}
    \caption{Inset figures from Fig.~\ref{fig:tilts-total-ppd} showing priors and posteriors on the
    marginal density of the population-level spin tilt distribution 
    averaged over a window of
    $\pm 0.1$ around $\cos \tau = 0$ under the \textsc{Subpopulations} (left) and \textsc{Correlation} (right) models.
    Colors and linestyles match Fig.~\ref{fig:tilts-total-ppd}.
    }
    \label{fig:inset-with-prior}
\end{figure}

\begin{figure*}
    \centering
    \includegraphics[width=0.8\linewidth]{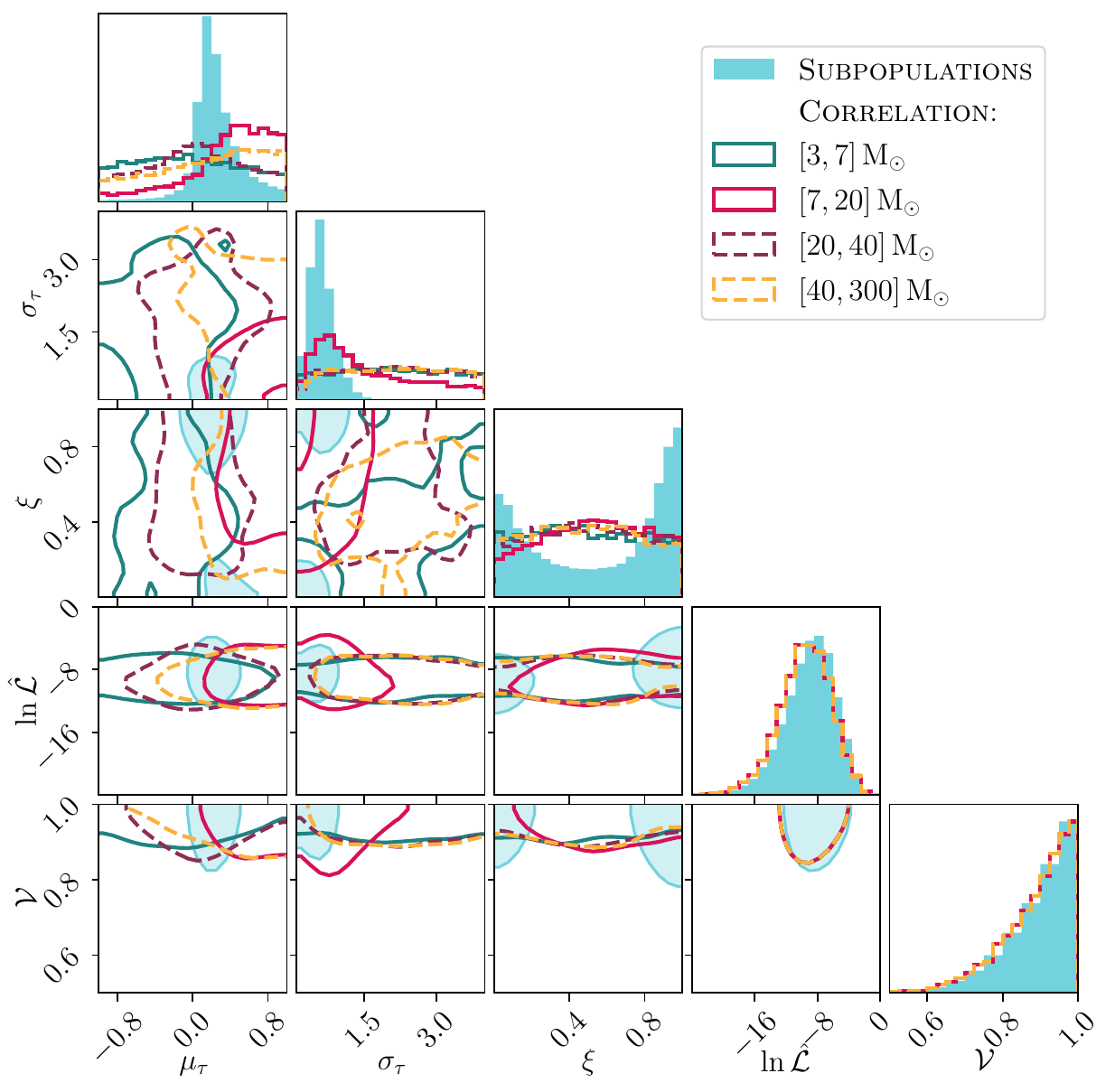}
    \caption{
    Marginal posteriors on the location $\mu_\tau$ and scale $\sigma_\tau$ of the truncated Gaussian in spin tilts,
    as well as the branching ratio $\xi$ of truncated Gaussian vs. isotropic tilt distribution.
    We also show the log-likelihood estimator $\ln \hat{\mathcal{L}}$ and variance $\mathcal{V}$ of that estimator.
    Log-likelihoods are shown relative to the maximum likelihood obtained under either model.
    Results shown are obtained under the \textsc{Subpopulations} model (blue, filled)
    as well as in each mass bin under the \textsc{Correlation} model (solid/dashed).
    For clarity, shading or lines only enclose the 50\% credible region.
    Note that $\sigma_\tau$ appears better measured under the \textsc{Subpopulations} model as we used a truncated Gaussian prior on $\sigma_\tau$,
    whereas we used a uniform prior in the \textsc{Correlation} model
    (cf. Tab.~\ref{tab:spin-priors}).
    }
    \label{fig:mutau-sigtau-xitau}
\end{figure*}

\begin{figure}
    \centering
    \includegraphics[width=0.5\linewidth]{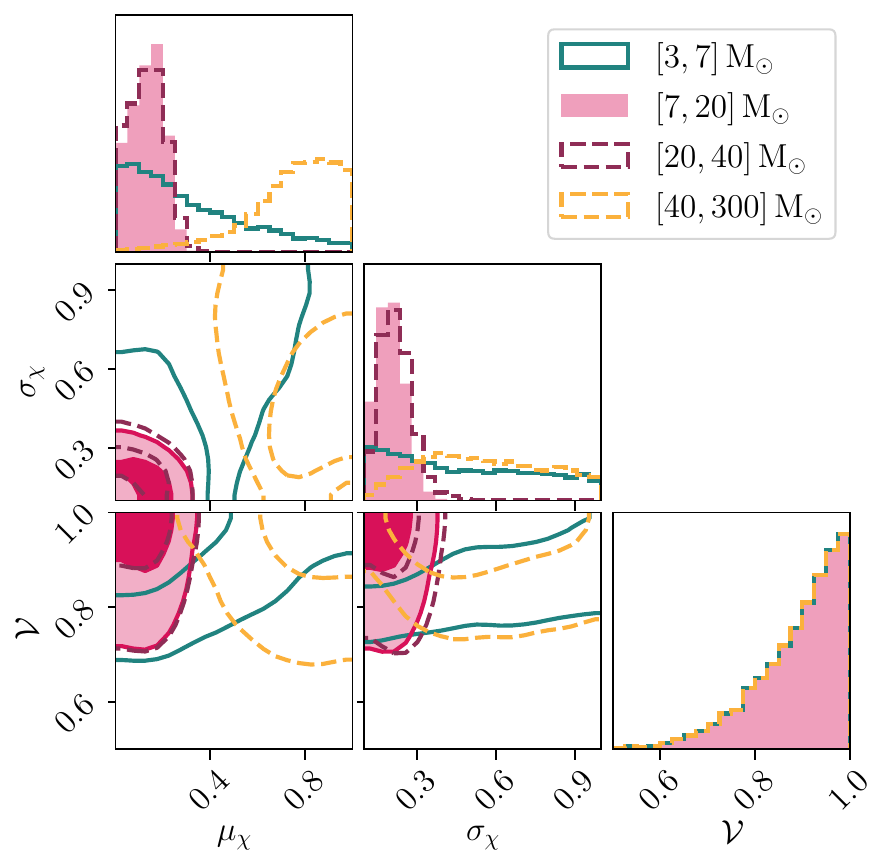}
    \caption{Marginal posteriors on the location $\mu_\chi$ and scale $\sigma_\chi$ of the spin magnitude distribution within each mass bin under the \textsc{Correlation} model,
    along with the variance $\mathcal{V}$ of the log-likelihood estimator.
    Contours enclose the 50\% and 90\% credible regions.
    }
    \label{fig:binned-muchi-sigchi}
\end{figure}

\begin{figure*}
    \centering
    \includegraphics[width=0.98\linewidth]{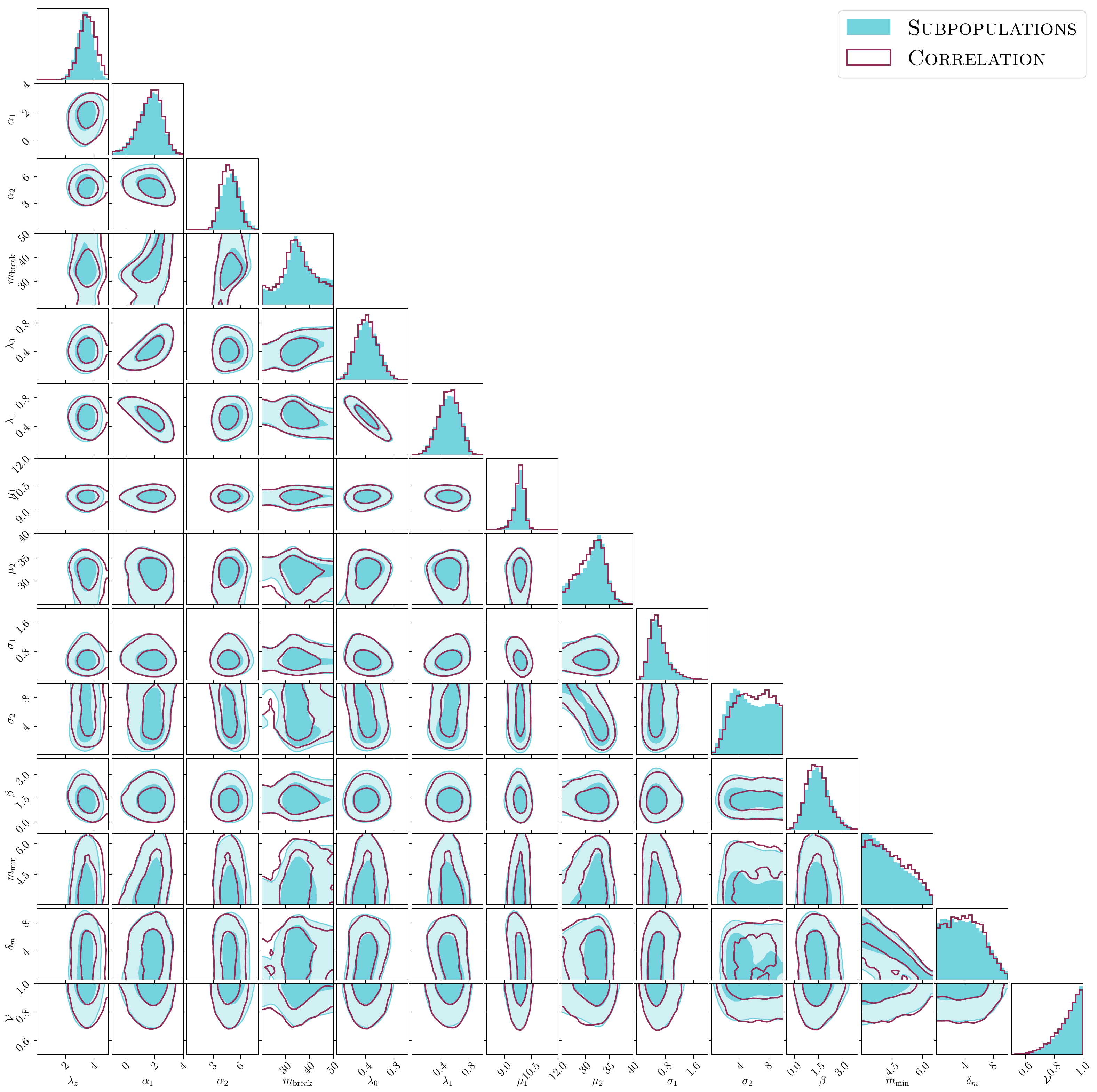}
    \caption{Marginal posteriors on the parameters of the primary mass, mass ratio, and redshift population distributions
    obtained under the \textsc{Subpopulations} model (blue, filled)
    as well as the \textsc{Correlation} model (purple).
    See Tab.~\ref{tab:mass-priors} for parameter definitions and priors.
    We also show the variance $\mathcal{V}$ of the log-likelihood estimator.
    Contours enclose the 50\% and 90\% credible regions.
    }
    \label{fig:mass-redshift}
\end{figure*}

In Fig.~\ref{fig:inset-with-prior}, we reproduce the inset figures from the top row of Fig.~\ref{fig:tilts-total-ppd} with their priors.
In Fig.~\ref{fig:mass-redshift},
we show
show marginal posteriors on the parameters of the redshift and mass distributions under the \textsc{Subpopulations} and \textsc{Correlation} models.
In Figs.~\ref{fig:mutau-sigtau-xitau} and \ref{fig:binned-muchi-sigchi}, we show marginal posteriors on population-level spin parameters
not presented elsewhere in this work.

\clearpage

\bibliography{refs}{}
\bibliographystyle{aasjournalv7}

\end{document}